\documentclass[aps,prl,preprint,tightenlines,superscriptaddress,showpacs,byrevtex]{revtex4}
\usepackage{graphicx} 
\usepackage{dcolumn}  

\graphicspath{{ps}}

\begin{document}

\vspace*{-3\baselineskip}
\resizebox{!}{3cm}{\includegraphics{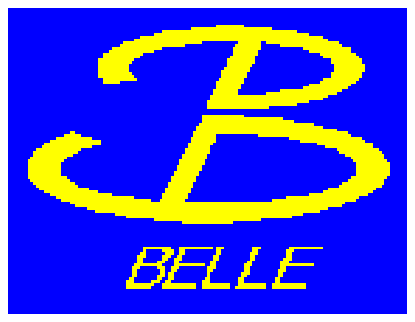}}

\preprint{\vbox{ \hbox{   }
                 \hbox{BELLE-CONF-0516}
                 \hbox{LP2005-PAPER-151}
                 \hbox{EPS05-488} 
}}

\title{ \quad\\[0.5cm]  Observation of $D_{s1}(2536)^+ \to D^+ \pi^- K^+$}



\affiliation{Aomori University, Aomori}
\affiliation{Budker Institute of Nuclear Physics, Novosibirsk}
\affiliation{Chiba University, Chiba}
\affiliation{Chonnam National University, Kwangju}
\affiliation{University of Cincinnati, Cincinnati, Ohio 45221}
\affiliation{University of Frankfurt, Frankfurt}
\affiliation{Gyeongsang National University, Chinju}
\affiliation{University of Hawaii, Honolulu, Hawaii 96822}
\affiliation{High Energy Accelerator Research Organization (KEK), Tsukuba}
\affiliation{Hiroshima Institute of Technology, Hiroshima}
\affiliation{Institute of High Energy Physics, Chinese Academy of Sciences, Beijing}
\affiliation{Institute of High Energy Physics, Vienna}
\affiliation{Institute for Theoretical and Experimental Physics, Moscow}
\affiliation{J. Stefan Institute, Ljubljana}
\affiliation{Kanagawa University, Yokohama}
\affiliation{Korea University, Seoul}
\affiliation{Kyoto University, Kyoto}
\affiliation{Kyungpook National University, Taegu}
\affiliation{Swiss Federal Institute of Technology of Lausanne, EPFL, Lausanne}
\affiliation{University of Ljubljana, Ljubljana}
\affiliation{University of Maribor, Maribor}
\affiliation{University of Melbourne, Victoria}
\affiliation{Nagoya University, Nagoya}
\affiliation{Nara Women's University, Nara}
\affiliation{National Central University, Chung-li}
\affiliation{National Kaohsiung Normal University, Kaohsiung}
\affiliation{National United University, Miao Li}
\affiliation{Department of Physics, National Taiwan University, Taipei}
\affiliation{H. Niewodniczanski Institute of Nuclear Physics, Krakow}
\affiliation{Nippon Dental University, Niigata}
\affiliation{Niigata University, Niigata}
\affiliation{Nova Gorica Polytechnic, Nova Gorica}
\affiliation{Osaka City University, Osaka}
\affiliation{Osaka University, Osaka}
\affiliation{Panjab University, Chandigarh}
\affiliation{Peking University, Beijing}
\affiliation{Princeton University, Princeton, New Jersey 08544}
\affiliation{RIKEN BNL Research Center, Upton, New York 11973}
\affiliation{Saga University, Saga}
\affiliation{University of Science and Technology of China, Hefei}
\affiliation{Seoul National University, Seoul}
\affiliation{Shinshu University, Nagano}
\affiliation{Sungkyunkwan University, Suwon}
\affiliation{University of Sydney, Sydney NSW}
\affiliation{Tata Institute of Fundamental Research, Bombay}
\affiliation{Toho University, Funabashi}
\affiliation{Tohoku Gakuin University, Tagajo}
\affiliation{Tohoku University, Sendai}
\affiliation{Department of Physics, University of Tokyo, Tokyo}
\affiliation{Tokyo Institute of Technology, Tokyo}
\affiliation{Tokyo Metropolitan University, Tokyo}
\affiliation{Tokyo University of Agriculture and Technology, Tokyo}
\affiliation{Toyama National College of Maritime Technology, Toyama}
\affiliation{University of Tsukuba, Tsukuba}
\affiliation{Utkal University, Bhubaneswer}
\affiliation{Virginia Polytechnic Institute and State University, Blacksburg, Virginia 24061}
\affiliation{Yonsei University, Seoul}
  \author{K.~Abe}\affiliation{High Energy Accelerator Research Organization (KEK), Tsukuba} 
  \author{K.~Abe}\affiliation{Tohoku Gakuin University, Tagajo} 
  \author{I.~Adachi}\affiliation{High Energy Accelerator Research Organization (KEK), Tsukuba} 
  \author{H.~Aihara}\affiliation{Department of Physics, University of Tokyo, Tokyo} 
  \author{K.~Aoki}\affiliation{Nagoya University, Nagoya} 
  \author{K.~Arinstein}\affiliation{Budker Institute of Nuclear Physics, Novosibirsk} 
  \author{Y.~Asano}\affiliation{University of Tsukuba, Tsukuba} 
  \author{T.~Aso}\affiliation{Toyama National College of Maritime Technology, Toyama} 
  \author{V.~Aulchenko}\affiliation{Budker Institute of Nuclear Physics, Novosibirsk} 
  \author{T.~Aushev}\affiliation{Institute for Theoretical and Experimental Physics, Moscow} 
% ...
\collaboration{The Belle Collaboration}

\begin{abstract}
We report the observation of the decay $D_{s1}(2536)^+ \to D^+ \pi^- K^+$.
We also measure the helicity angle distributions in the decay 
$D_{s1}(2536)\to D^{*+} K^0_S$ and thus constrain the contributions and the phase difference of $D$ and $S$ wave amplitudes in this decay. 
The results are based on a 281~fb$^{-1}$ data sample collected with the Belle detector near the $\Upsilon(4S)$ resonance, at the KEKB asymmetric energy 
$e^+ e^-$ collider.
\end{abstract}


\maketitle

\tighten

{\renewcommand{\thefootnote}{\fnsymbol{footnote}}}
\setcounter{footnote}{0}

\section{Introduction}

Two states $D_{sJ}(2317)^+$ and $D_{sJ}(2460)^+$ have been discovered recently both in continuum $e^+e^-$ annihilation near $\sqrt{s}=10.6$ GeV/c$^2$
 and in $B$ meson decays~\cite{dsj},~\cite{dspipi}.
Their decay properties are consistent with the assumption that these are $J^P=0^+,1^+$ states with $j=L+S_{\bar{s}}=1/2$. Here
$L=1$ is the orbital momentum, $S_{\bar{s}}$ is the spin of the light antiquark. However, their masses are unexpectedly low~\cite{dsj_theory}.
This has renewed interest in measurements of P-wave excited charm mesons.

We report the first observation of the decay $D_{s1}(2536)^+\to D^+\pi^-K^+$ (the inclusion of charge conjugate modes
is implied throughout the paper). The
$D^+\pi^-$ pair in the final state is the only $D \pi$ combination that cannot come from a $D^*$ resonance. Note that $D^{*0}$ mesons can only be produced
virtually here since $M_{D^{*0}} < M_{D^+} + M_{\pi^-}$. 
The $D_{s1}(2536)^+ \to D^+\pi^-K^+$ and 
$D_{s1}(2536)^+\to D_s^+\pi^+\pi^-$~\cite{dspipi} modes are the only known three-body decays of the $D_{s1}(2536)^+$. 

In addition, we have performed an angular analysis of the $D_{s1}(2536)^+\to D^{*+} K^0_S$ mode. 
In the limit of infinite $c$ quark mass this decay of a $J^P=1^+$, $j=3/2$ 
state should proceed via a pure D-wave~\cite{HQET}.
The corresponding decay of its partner, the $D_{sJ}(2460)^+$, which is believed to be a $1^+$, $j=1/2$ state is energetically forbidden, but if it were
allowed it would proceed via a pure S wave. 
Since heavy quark symmetry is not exact, the two $1^+$ states can mix with each other. In particular, an S wave component can 
appear in the decay $D_{s1}(2536)^+\to D^* K$. 
Moreover, even if the mixing is small, the S wave component can give a sizeable contribution to the width because
the D wave contribution is strongly suppressed by the small energy release in the $D_{s1}(2536)^+\to D^* K$ decay.
The angular decomposition in S and D waves for the analogous decays of the $1^+$, $j=3/2$\ \ $D_1(2420)^{0,+}$ mesons to $D^{*+}\pi^-$, $D^{*0}\pi^+$
was performed more than 10 years ago by CLEO~\cite{cleo}, 
but currently no results on the $D_{s1}(2536)^+$ exist.

This study is based on a data sample of $253
\,\mathrm{fb}^{-1}$ collected at the $\Upsilon(4S)$ resonance and
$28\,\mathrm{fb}^{-1}$ at an energy $60\,{\mathrm{MeV}}$ below the
resonance with the Belle
detector
at the KEKB asymmetric-energy
$e^+e^-$ (3.5 on 8~GeV) collider~\cite{KEKB}.
The Belle detector is a large-solid-angle magnetic
spectrometer that
consists of a silicon vertex detector (SVD),
a 50-layer central drift chamber (CDC), an array of
aerogel threshold \v{C}erenkov counters (ACC), 
a barrel-like arrangement of time-of-flight
scintillation counters (TOF), and an electromagnetic calorimeter
comprised of CsI(Tl) crystals (ECL) located inside 
a super-conducting solenoid coil that provides a 1.5~T
magnetic field.  An iron flux-return located outside of
the coil is instrumented to detect $K_L^0$ mesons and to identify
muons (KLM).  The detector
is described in detail elsewhere~\cite{Belle}.
Two inner detector configurations were used. A 2.0 cm beampipe
and a 3-layer silicon vertex detector was used for the first sample
of $155\,\mathrm{fb}^{-1}$, while a 1.5 cm beampipe, a 4-layer
silicon detector and a small-cell inner drift chamber were used to record  
the remaining $126\,\mathrm{fb}^{-1}$~\cite{Ushiroda}.  

In Monte Carlo, $D_{s1}(2536)^+$ from $e^+e^-$ annihilation, particle decays and the detailed detector response is simulated using the
{\tt PYTHIA}, {\tt EvtGen} and {\tt GEANT} packages~\cite{mcpackages} respectively.
The $D^0$ and $D^+$ decay modes used in reconstruction are generated with their resonant substructures taken from the PDG~\cite{pdg} but
neglecting any interference effects. 
The width of the $D_{s1}(2536)^+$ resonance in the simulation is set to zero.
Only the D wave matrix element is used for the $D_{s1}(2536)^+\to D^{*+}K^0_S$ decay.
As is shown below no clear resonant substructure is visible in the decay $D_{s1}(2536)^+\to D^+\pi^-K^+$. 
Therefore, it is simulated as a three-body phase space decay. 

\section{$D_{s1}(2536)^+\to D^+\pi^-K^+$ decay and calculation of $\frac{\mathcal{B}(D_{s1}^+\to D^+\pi^- K^+)}{\mathcal{B}(D_{s1}^+\to D^{*+}K^0)}$.}

$\pi^{\pm}$ and $K^{\pm}$ candidates are required to originate from the
vicinity of the event dependent interaction point.
To identify kaons, the $dE/dx$, time of flight and \v{C}erenkov light yield
information for each track are combined to form kaon $\mathcal{L}_K$ and pion $\mathcal{L}_{\pi}$ likelihoods and the requirement
$\mathcal{L}_K/(\mathcal{L}_K+\mathcal{L}_{\pi}) > 0.1$ is imposed.
$K^0_S$ candidates are reconstructed via the $\pi^+ \pi^-$ decay channel.
$D^0$ and $D^+$ mesons are reconstructed using $K^-\pi^+$, $K^0_S \pi^+\pi^-$, $K^-\pi^+\pi^+\pi^-$ and 
$K^0_S \pi^+$, $K^-\pi^+\pi^+$ decay modes, respectively.
All combinations with masses within $\pm20$~MeV/c$^2$ of the nominal $D$ mass are selected and then a mass and vertex
constrained fit is applied.

Candidate $D^{*+}$'s are reconstructed using the $D^0\pi^+$ mode. The slow $\pi^+$ momentum resolution suffers from multiple scattering. 
It is improved by a track refit procedure in which the $\pi^+$ origin point is constrained by the intersection of the $D$ momentum 
and the known region of $e^+e^-$ interactions. The same procedure is applied to the slow $\pi^-$ and $K^+$ from the 
$D_{s1}(2536)^+\to D^+\pi^-K^+$ decay.
The $M_{D^0\pi^+}$ mass is required to be within $\pm 1.5$~MeV/c$^2$ around the $D^{*+}$ nominal value.
The $D^{*+}$ mass constraint fit is not applied. Instead, the mass difference $M_{D^0\pi^+K^0_S} - M_{D^0\pi^+}$ is used for $D_{s1}(2536)^+$ where the 
errors in $D^{*+}$ momentum essentially cancel out.

It is known that the momentum spectrum of the excited charm resonances from continuum $e^+e^-$ annihilation is hard. 
In addition, due to the strong magnetic field in BELLE
the reconstruction efficiency for slow $\pi^{\pm}$ and $K^+$ mesons is larger for higher $D_{s1}(2536)^+$ momenta.
Therefore, to reduce the combinatorial background, it is required that $x_P > 0.8$, where the scaled momentum $x_P$ is defined as 
$x_P = p^\ast/p^\ast_{\text{max}}$. Here $p^\ast$ is the momentum 
of the $D_{s1}(2536)^+$ candidate in the $e^+e^-$ center of mass frame. $p^\ast_{\text{max}}$ is the momentum which the candidate would have
if it carried all of the beam energy $E^\ast_{\rm beam}$ in the same frame:
$p^\ast_{\text{max}} = \sqrt{E^{\ast 2}_{\rm beam}-M^2}$.

\begin{figure}[htb]
\includegraphics[width=0.54\textwidth]{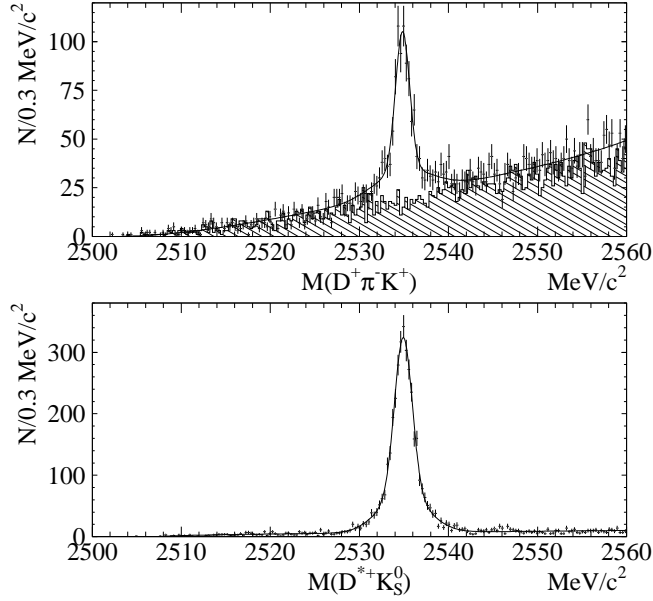}
\caption{$D_{s1}(2536)^+$ mass spectra for $D^+\pi^-K^+$ (top) and $D^{*+}K^0_S$ (bottom) decay modes. The latter is calculated using the mass 
difference $(M_{D^0\pi^+K^0_S} - M_{D^0\pi^+}) + M^{\rm{PDG}}_{D^{*+}}$. 
The hatched histogram in the top plot shows the corresponding spectrum of wrong sign $D^+\pi^+K^-$ combinations.
The fit is described in the text. The fit results are listed in Table~\ref{fit}.}
\label{res}
\end{figure}

The mass $M_{D^+\pi^-K^+}$ and  the mass difference $(M_{D^0\pi^+K^0_S} - M_{D^0\pi^+}) + M^{\rm{PDG}}_{D^{*+}}$ for all
accepted combinations are plotted in Fig.~\ref{res}. Here and in the following the PDG superscript denotes the nominal mass value from~\cite{pdg}.
A clear peak for the new decay channel $D_{s1}(2536)^+\to D^+\pi^-K^+$ is visible in Fig.~\ref{res}.
The mass spectrum of the wrong sign combinations $D^+\pi^+K^-$ shown by the hatched histogram has no enhancement in the $D_{s1}(2536)^+$ region.

To calculate the number of $D_{s1}(2536)^+$ decays, the distributions in Fig.~\ref{res} are fit
to the sum of two Gaussians. Their central values are required to be equal.
To ensure that the second Gaussian is always wider than the first one, its width is chosen to be of the form 
$\sigma_2=\sqrt{\sigma_1^2+\Delta\sigma^2}$.
The position of the peak, $\sigma_1$, $\Delta\sigma$, the fraction of events in the first Gaussian and the total number of events in two Gaussians are allowed
to vary in the fit.
The background for the three-body $D^+\pi^-K^+$ mode is parameterized by the second order polynomial multiplied by the 
function $(M-M^{threshold}_{D^+\pi^-K^+})^2$, where $M^{threshold}_{D^+\pi^-K^+} = M_{D^+}^{PDG} + M_{\pi^-}^{PDG} + M_{K^+}^{PDG}$. 
For the two-body $D^{*+}K^0_S$ mode the background parametrization is chosen to be of the form $\sqrt{M-M^{threshold}_{D^{*+}K^0_S}}$ 
times a first order polynomial, where 
$M^{threshold}_{D^{*+}K^0_S} = M_{D^{*+}}^{PDG} + M_{K^0_S}^{PDG}$.
Table~\ref{fit} contains the fit results together with the parameters of the Gaussians obtained from Monte Carlo. 
There is a small fraction of events that contribute two entries to the $D_{s1}(2536)^+$ signal region in the mass plot. 
The last row in the Table~\ref{fit} shows the excess of such events in comparison with the same number averaged over the left and the right 
sideband.
The signal and the sidebands are defined as 
$|\Delta M_{D_{s1}^+}|<5$ MeV/c$^2$, $10$ MeV/c$^2$ $<|\Delta M_{D_{s1}^+}|<20$ MeV/c$^2$,
respectively, where $\Delta M_{D_{s1}^+}$ is measured relative to the peak position obtained from the fit to
Fig.~\ref{res}.

\begin{table}[htb]
\caption{ Fit results of the spectra shown in Fig.~\ref{res}: 
number of events in two Gaussians, fraction of events in the narrower 1st Gaussian, $\sigma_1$, 
$\Delta\sigma$ and $M_{D_{s1}}-M_{D_{s1}}^{PDG}$. The last three values 
are given in MeV/c$^2$.
Note that the  value of $M_{D_{s1}}^{PDG}$ is known with errors of $\pm0.34\pm0.5$ MeV/c$^2$.
The sigma of the 2nd Gaussian is $\sigma_2=\sqrt{\sigma_1^2+\Delta\sigma^2}$. The 3rd and 5th columns contain the corresponding Gaussian parameters 
obtained in Monte Carlo. 
The number of double counted events after sideband subtraction
is given in the last row.}
\label{fit}
\begin{tabular}
{|@{\hspace{0.3cm}}l@{\hspace{0.3cm}}|@{\hspace{0.3cm}}c@{\hspace{0.3cm}}|@{\hspace{0.3cm}}c@{\hspace{0.3cm}}||@{\hspace{0.3cm}}c@{\hspace{0.3cm}}|
@{\hspace{0.3cm}}c@{\hspace{0.3cm}}|}
\hline
                                         & $(D^+\pi^-K^+)_{Data}$& $(D^+\pi^-K^+)_{MC}$& $(D^{*+}K^0_S)_{Data}$& $(D^{*+}K^0_S)_{MC}$ \\ \hline
N events                                 & $\bf { 802 \pm  56}$  &                     & ${\bf  3474 \pm  64}$ &                      \\ \hline
1st G. fraction                          & ${\bf  0.58\pm0.07}$  & $0.769\pm0.014$     & ${\bf  0.60 \pm0.04}$ & $0.862\pm0.013$      \\ \hline
$\sigma_1$                               & ${\bf  0.79\pm0.07}$  & $0.478\pm0.011$     & ${\bf  0.98 \pm0.04}$ & $0.746\pm0.014$      \\ \hline
$\Delta\sigma$                           & ${\bf  3.1 \pm0.6 }$  & $1.85 \pm0.11 $     & ${\bf  2.46 \pm0.15}$ & $2.49  \pm0.18$      \\ \hline
$M_{D_{s1}}-M_{D_{s1}}^{PDG}$            & ${\bf -0.51\pm0.06}$  & $0.014\pm0.008$     & ${\bf -0.42 \pm0.03}$ & $0.023\pm0.012$      \\ \hline
double counting                          & {\bf $28-\frac{20}{2}=18$}&                     & {\bf $105-\frac{18}{2}=96$} &            \\ \hline
\end{tabular}
\end{table}

\begin{figure}[htb]
\includegraphics[width=0.54\textwidth]{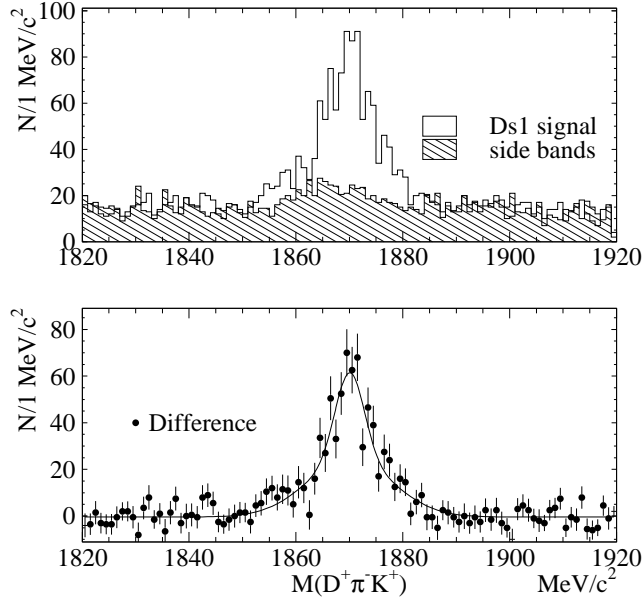}
\caption{$D^+$ mass spectrum for the $D_{s1}(2536)^+$ signal band ($|\Delta M_{D^+\pi^-K^+}|<5$~MeV/c$^2$, open histogram in the top plot) 
and the sidebands ($10<|\Delta M_{D^+\pi^-K^+}|<20$~MeV/c$^2$, normalized to the signal interval, hatched histogram).
Here $\Delta M_{D^+\pi^-K^+}=M_{D^+\pi^-K^+}-M_{D_{s1}}^0$, 
$M_{D_{s1}}^0$ is the $D_{s1}(2536)^+$ peak position in the top plot of Fig.~\ref{res}. 
The bottom plot shows the difference. The solid curve shows the results of the fit described in the text.}
\label{dp}
\end{figure}

To cross-check the results, the $D^+$ mass spectrum is plotted in Fig.~\ref{dp} for the $D_{s1}(2536)^+$ signal
and sidebands. The latter is normalized to the area of the signal interval.
The sideband subtracted plot shown in the bottom of Fig.~\ref{dp} is fit 
to a double Gaussian as above and a constant background.
The resulting yield $739 \pm  51$ is consistent with the yield
$679\pm48$ obtained from the fit of the $D_{s1}(2536)^+$ mass spectrum. 
The constant background level is found to be $-0.4 \pm 0.6$, which is consistent with zero.
The enhancement in the $D^+$ mass region observed 
in the $D_{s1}(2536)^+$ sidebands (top plot of Fig.~\ref{dp}) is due to 
combinations of a real $D^+$ with a random $\pi^- K^+$ pair in the event. 

Due to the low momenta of the final state particles, the $D_{s1}(2536)^+$ reconstruction efficiency strongly depends on its momentum,
which is found to be harder in data than in Monte Carlo.
%
%
%
Therefore 
the ratio of $D_{s1}(2536)^+$ branching fractions is calculated
using the following formula:
$$\frac{\mathcal{B}(D_{s1}(2536)^+\to D^+\pi^-K^+)}{\mathcal{B}(D_{s1}(2536)^+\to D^{*+}K^0)} = $$
$$\frac{(N_{D\pi K}-n_{D\pi K}^{\rm \bf dbl\ cnt})}
{
(N_{D^{*+}K^0_S}-n_{D^{*+}K^0_S}^{\rm \bf dbl\ cnt})
}
\left[\frac{\sum_{p_i}{N_{D^{*+}K^0_S}^{p_i}\frac{\epsilon_{D^+\pi^-K^+}^{p_i}}{\epsilon_{D^{*+}K^0_S}^{p_i}}}}
     {\sum_{p_i}{N_{D^{*+}K^0_S}^{p_i}}}\right]^{-1}
\!\!
\frac{\mathcal{B}(D^{*+}\!\!\to\! D^0\pi^+) \mathcal{B}(D^0) \mathcal{B}(K^0\!\! \to\! K^0_S\! \to\! \pi^+\pi^-)}
{\mathcal{B}(D^+)}.$$
Here, $\epsilon_{D^+\pi^-K^+}^{p_i}$ and $\epsilon_{D^{*+}K^0_S}^{p_i}$ are the reconstruction efficiencies
in individual momentum bins of $D_{s1}(2536)^+$, $N_{D^{*+}K^0_S}^{p_i}$ are the
number of $D_{s1}(2536)^+\to D^{*+}K^0_S$ decays
observed in a
given momentum bin,
and the sum runs over the momentum bins with $x_P>0.8$.
$N_{D\pi K,\ D^{*+}K^0_S}$ and $n_{D\pi K,\ D^{*+}K^0_S}^{\rm \bf dbl\ cnt}$
are the total number of decays obtained by performing the fit to the $D_{s1}(2536)^+$ mass
spectra and the number of double counted events, respectively (see Table~\ref{fit}).
$\mathcal{B}(D^0,\ D^+)$ is the sum of branching fractions of $D^0$ and $D^+$ modes used in the reconstruction~\cite{pdg}. 
For modes with $K^0_S$'s in the final state the branching fractions are in addition multiplied by
$\mathcal{B}(K^0 \to K^0_S \to \pi^+\pi^-)=\frac{1}{2}\cdot \mathcal{B}(K^0 \to \pi^+\pi^-)$. 
The dependence of the efficiency on the initial
polarization of the $D_{s1}(2536)^+$ is checked and found to be
negligible. 
The efficiency does not
depend on the $D^{*+}$ helicity angle in the $D^{*+}K^0_S$ decay and thus on the proportions of S and D waves.
For the $D^+\pi^-K^+$ mode the efficiency is independent
of the $D^+\pi^-$, $K^+\pi^-$ masses and the $D^+\pi^-$ helicity angle.
The ratio of branching fractions is found to be
$$\frac{\mathcal{B}(D_{s1}(2536)^+\to D^+\pi^-K^+)}{\mathcal{B}(D_{s1}(2536)^+\to D^{*+}K^0)} = 
(2.8\pm 0.2\pm0.4)\%.$$
%
The systematic error receives contribution from different sources listed in
Table~\ref{syst}.
A possible difference between the data and Mone Carlo in evaluation of the
tracking efficiency was estimated using partially reconstructed $D^{*+}$
decays. The tracking errors due to slow $K^+$, $\pi^{\pm}$ and pions from $K^0_S$ are added
linearly. The uncertainty in the kaon particle identification is estimated using $D^{*+}$ decays.
Uncertainty in the ratio of $D^+$ and $D^0$ efficiencies was determined by a
comparison of different decay modes used in the reconstruction. The
largest contribution to the systematic uncertainty arises due to the
fitting model and was evaluated by comparing the fit results using
different binnings of the $D^+\pi^-K^+$ and $D^{*+}K^0_S$ mass spectra. The largest discrepancy
(in the $D^+\pi^-K^+$ decay mode) is taken as the systematic error. 
Finally, the
contribution due to the assumption that the efficiency does not depend on
the decay angles of the $D_{s1}(2536)^+$ and its initial polarization, $D\pi$ helicity angles
and $M(D^+\pi^-)$ is evaluated by comparing
the yields of events using either an average or differential efficiency
in the specified variables.
%
%
 The total systematic error is found to be
12\% (Table~\ref{syst}). 

\begin{table}[htb]
\caption{Systematic uncertainties for $\frac{\mathcal{B}(D_{s1}(2536)^+\to D^+\pi^-K^+)}{\mathcal{B}(D_{s1}(2536)^+\to D^{*+}K^0)}$.}
\label{syst}
\begin{tabular}
{ @{\hspace{0.3cm}}l@{\hspace{0.3cm}} @{\hspace{0.3cm}}c@{\hspace{0.3cm}} }
\hline\hline
Source & Uncertainty, \% \\ \hline
Slow $\pi^{\pm}$ tracking efficiency & 1.5 \\
Slow $K^+$ tracking efficiency       & 1   \\
Slow $K^0_S$ tracking efficiency     & 5   \\
Slow $K^+$ particle identification   & 1.2 \\
Ratio of $D^+$ and $D^0$ efficiencies& 3.5 \\
Fit to $M(D^+\pi^-K^+)$ distribution & 8.5 \\
Efficiency independence on $D_{s1}(2536)^+$ polarization, & \\
$D\pi$ helicity angle and $M(D^+\pi^-)$ & 1.1 \\ \hline
Total                                & 12.0\\ \hline\hline
\end{tabular}
\end{table}

\begin{figure}[htb]
\includegraphics[width=0.54\textwidth]{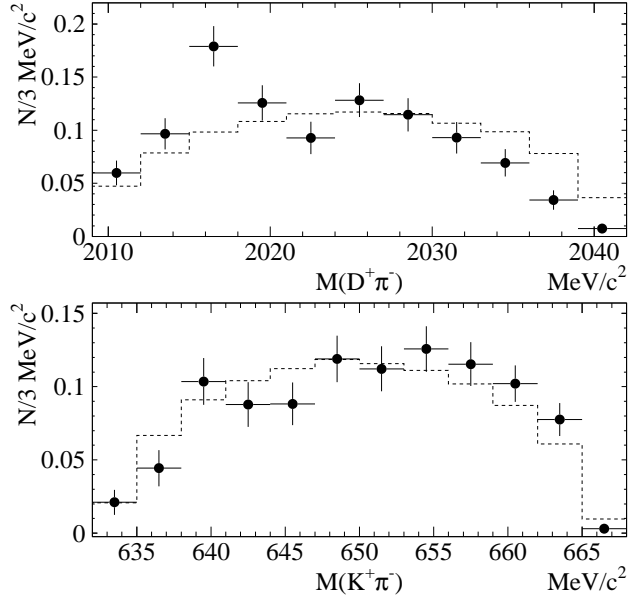}
\caption{Normalized mass spectra of $D^+\pi^-$ (top) and $K^+\pi^-$ (bottom) pairs 
from $D_{s1}(2536)^+\to D^+\pi^-K^+$ decay obtained from fits to the $D^+\pi^-K^+$ mass distributions 
in different $D^+\pi^-$ or $K^+\pi^-$ mass bins.
The dashed histograms show the corresponding distributions for the $D_{s1}(2536)^+$ phase space decay in Monte Carlo.}
\label{mdpi_mkpi}
\end{figure}

The $D^+\pi^-$ and $K^+\pi^-$ mass distributions for the 
$D_{s1}(2536)^+\to D^+\pi^-K^+$ decay are shown in Fig.~\ref{mdpi_mkpi}.
The $D_{s1}(2536)^+$ signal yield is obtained from fits to the $D^+\pi^-K^+$ mass 
distribution in bins of $D^+\pi^-$ and $K^+\pi^-$ mass. 
All Gaussian parameters except the total
number of events are fixed in the fit to the values obtained from Fig.~\ref{res}. 
The position of the threshold used for the background description 
depends on the chosen bin.
The areas of the plots have been normalized to unity. The
plots are not efficiency corrected since the differential efficiency does not depend on 
$D^+\pi^-$ or $K^+\pi^-$ masses to within the errors of Monte Carlo statistics. 
No dominant resonant substructure is visible in Fig.~\ref{mdpi_mkpi}.

\section{Angular analysis of $D_{s1}(2536)^+\to D^{*+}K^0_S$ decay.}

\begin{figure}[htb]
\includegraphics[width=0.54\textwidth]{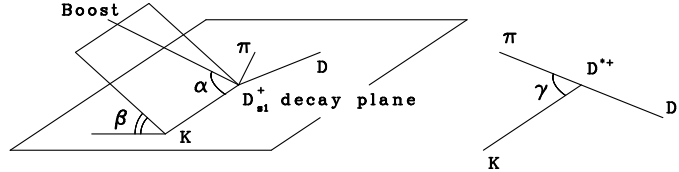}
\caption{Definitions of the angles $\alpha$, $\beta$ and $\gamma$. 
The first two are measured in the $D_{s1}(2536)^+$ rest frame, the last one -- in the $D^{*+}$ frame. 
``Boost'' changes the $e^+e^-$ center of mass system to the $D_{s1}(2536)^+$ frame.}
\label{angles_scheme}
\end{figure}

The $D_{s1}(2536)^+\to D^{*+}K^0_S$ decay can be described by three angles $\alpha$, $\beta$ and $\gamma$ defined as shown in Fig.~\ref{angles_scheme}.
The angles $\alpha$ and $\beta$ are measured in the $D_{s1}^+$ rest frame with respect to the direction of the boost needed to go from the $e^+e^-$ center of mass 
frame to the $D_{s1}^+$ rest frame.
$\alpha$ is the angle between the boost direction and the $K^0_S$ momentum. $\beta$ is the angle between the decay plane and 
the plane formed by the $K^0_S$ and the boost direction. The third angle $\gamma$ is defined in the
$D^{*+}$ rest frame between $\pi^+$ and $K^0_S$. 

\begin{figure}[htb]
\includegraphics[width=0.54\textwidth]{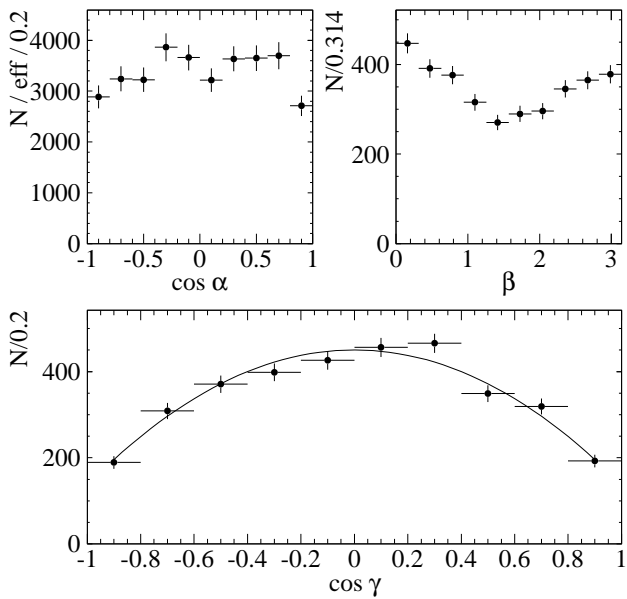}
\caption{Angular distributions for the $D_{s1}(2536)^+\to D^{*+}K^0_S$ channel. 
The definitions of angles $\alpha$, $\beta$ and $\gamma$ are given in the text and in Fig.~\ref{angles_scheme}. 
$D_{s1}(2536)^+$ reconstruction efficiency depends slightly on $\cos\alpha$, therefore the upper left plot is efficiency corrected. 
A fit to the $\cos\gamma$ distribution is described in the text.}
\label{angles}
\end{figure}

The measured $\cos\alpha$, $\beta$ and $\cos\gamma$ angular distributions are shown in Fig.~\ref{angles}. 
They represent the signal yield obtained from fits to the $D^{*+}K^0_S$ mass
spectra in bins of individual angular variable.
All Gaussian parameters except the normalization are fixed 
in the fit to the values obtained from the overall spectrum in Fig.~\ref{res}. 
The first distribution is efficiency corrected
since there is a slight linear efficiency dependence on $\cos\alpha$:
the relative difference between the values at $-1$ and at $+1$ is about 12\%.
Since the differential distribution in $\cos\alpha$ includes only even
powers of $\cos\alpha$, such a linear dependence
does not induce any biases when integrating over the whole allowed
interval $[-1,+1]$ and thus no dependence of the integrated efficiency on
the initial polarization of $D_{s1}(2536)^+$.
No significant efficiency dependence on $\beta$ and $\cos\gamma$ is observed
and therefore the corresponding two spectra in
Fig.~\ref{angles} are not efficiency corrected. 
One can see that the first two distributions are not flat. This means that $D_{s1}(2536)^+$ is produced polarized and that 
it does not decay in a pure S wave. The last distribution is more important. For a pure S or D wave decay it should either be flat 
or have the form $(1+3\cos^2\gamma)$, respectively.
In the general case of interference between S and D wave amplitudes it becomes a linear combination of $\sin^2\gamma$ and $\cos^2\gamma$:
$$ \frac{1}{N}\frac{dN}{d\, \cos\gamma} =
\frac{1}{2}\{ R + (1-R)(\frac{1+3\cos^2\gamma}{2}) + \sqrt{2R(1-R)}\cos\phi (1-3\cos^2\gamma)\},$$
where $R=\Gamma_S/(\Gamma_S+\Gamma_D)$, $\Gamma_{S,D}$ are the S, D wave partial widths respectively, $\phi$ is the relative phase between 
the two amplitudes. This is similar to the case of $D^*_1(2420)^0\to D^{*+}\pi^-$ and $D^*_1(2420)^+\to D^{*0}\pi^+$ decays studied by CLEO~\cite{cleo}. 

\begin{figure}[htb]
\includegraphics[width=0.54\textwidth]{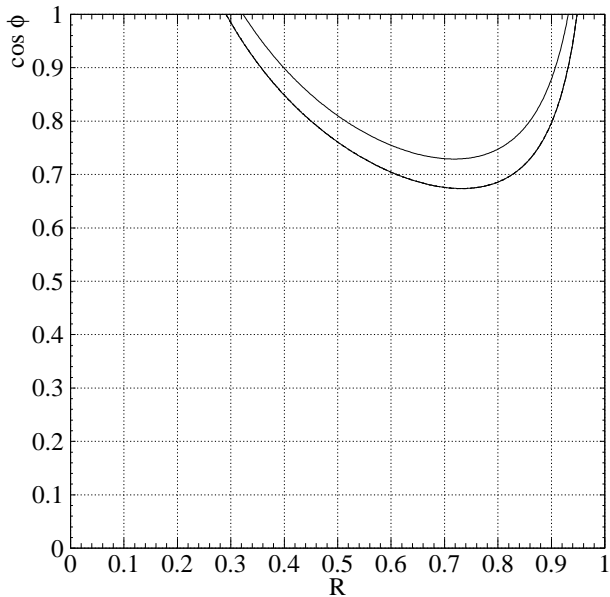}
\caption{Plot of cosine of the relative phase of S and D wave amplitudes in the $D_{s1}(2536)^+\to D^{*+}K^0_S$ decay versus 
$R=\Gamma_S/(\Gamma_S+\Gamma_D)$. The two curves bound the region that corresponds to a $\pm1\sigma$ deviation in the measured parameter $A$.}
\label{Rphi}
\end{figure}

Fitting the $\cos\gamma$ distribution in Fig.~\ref{angles} to the form $1+A\cos^2\gamma$ yields 
$A=-0.70\pm0.03$ and a $\chi^2$ per degree of freedom of $1.39$. 
It was checked that 
the $A$ parameter obtained on the subsamples defined by the cuts 
$\cos\alpha<0$, $\cos\alpha>0$, $\beta<\pi/2$, $\beta>\pi/2$, $x_P<0.93$ or $x_P>0.93$ agree with each other within statistical accuracy.
Knowledge of $A$ constrains the contribution of S wave to the width and the relative phase $\phi$:
$$ \cos\phi = \frac{\frac{3-A}{3+A}-R}{2\sqrt{2R(1-R)}}.$$
The allowed range $|\cos\phi|\le 1$ is shown in Fig.~\ref{Rphi}. Two lines in this plot bound the region
which corresponds to $\pm1\sigma$ deviation in $A$. 
Regardless of the value of
$\phi$, the S wave contribution is limited
from below and from above by the values corresponding to $\phi = 0$.
Conservatively taking the value of A to be 2$\sigma$ below 
the central value, the limits
obtained are:
$$0.277<R<0.955.$$
The corresponding limit for $\phi$ is: $|\phi|<42^{\circ}$.

In conclusion, a new decay channel $D_{s1}(2536)^+\to D^+\pi^-K^+$ is observed. The $D^+\pi^-$ pair is the only $D\pi$ combination that cannot come 
from a real $D^*$ resonance. It can be produced in $D_{s1}(2536)^+$ two-body decays only through the virtual resonances 
$D^{*0}$, broad $D^{*0}_0$ or $D^{*0}_2(2460)$. 
In addition, the $D^+\pi^-K^+$ final state can be formed by two-body decays to a $D^+$ and a virtual $K^{*0}$ or higher $K^*$ resonance. 
No clear resonant substructure is found in the $D^+\pi^-K^+$ system. 
The ratio of branching fractions  
$\frac{\mathcal{B}(D_{s1}(2536)^+\to D^+\pi^-K^+)}{\mathcal{B}(D_{s1}(2536)^+\to D^{*+}K^0)}$ is measured to be 
$(2.8\pm 0.2\pm0.4)\%$. 
An angular analysis of the normalization decay $D_{s1}(2536)^+\to D^{*+}K^0_S$ is also performed. 
The $D_{s1}(2536)^+$ may mix with another
$J^P=1^+$, $j=1/2$ state, which is presumably the recently discovered $D_{sJ}(2460)^+$ meson and can decay in an S wave.
Since the energy release in this reaction is small, the D wave is suppressed and the S wave can give a sizeable contribution to the total width
even if the mixing is small.
The measured $1-(0.70\pm0.03)\cos^2\gamma$ $D^{*+}$ helicity angular distribution 
constrains the relative
fraction of the S wave component to the range $0.277<R<0.955$, independent of the
phase $\phi$.

We thank the KEKB group for the excellent operation of the
accelerator, the KEK cryogenics group for the efficient
operation of the solenoid, and the KEK computer group and
the National Institute of Informatics for valuable computing
and Super-SINET network support. We acknowledge support from
the Ministry of Education, Culture, Sports, Science, and
Technology of Japan and the Japan Society for the Promotion
of Science; the Australian Research Council and the
Australian Department of Education, Science and Training;
the National Science Foundation of China under contract
No.~10175071; the Department of Science and Technology of
India; the BK21 program of the Ministry of Education of
Korea and the CHEP SRC program of the Korea Science and
Engineering Foundation; the Polish State Committee for
Scientific Research under contract No.~2P03B 01324; the
Ministry of Science and Technology of the Russian
Federation; the Ministry of Higher Education, 
Science and Technology of the Republic of Slovenia;  
the Swiss National Science Foundation; the National Science Council and
the Ministry of Education of Taiwan; and the U.S.\
Department of Energy.


\end{document}